\documentclass[a4paper,11pt]{article}
\pdfoutput=1 
\usepackage{jcappub} 
\usepackage[T1]{fontenc}

\usepackage{amsmath,amssymb,amsfonts,bm,mathtools}
\usepackage{graphicx}
\usepackage{booktabs}
\usepackage{tabularx}
\usepackage{float}
\usepackage{subcaption}
\usepackage{pifont}
\usepackage{enumerate}
\usepackage{microtype}  
\usepackage{xcolor}
\usepackage{hyperref}
\usepackage{natbib}
\usepackage{tikz}
\usepackage{tablefootnote}
\usepackage{afterpage}

\newcommand{\orcid}[1]{\href{https://orcid.org/#1}{\,\includegraphics[width=8px]{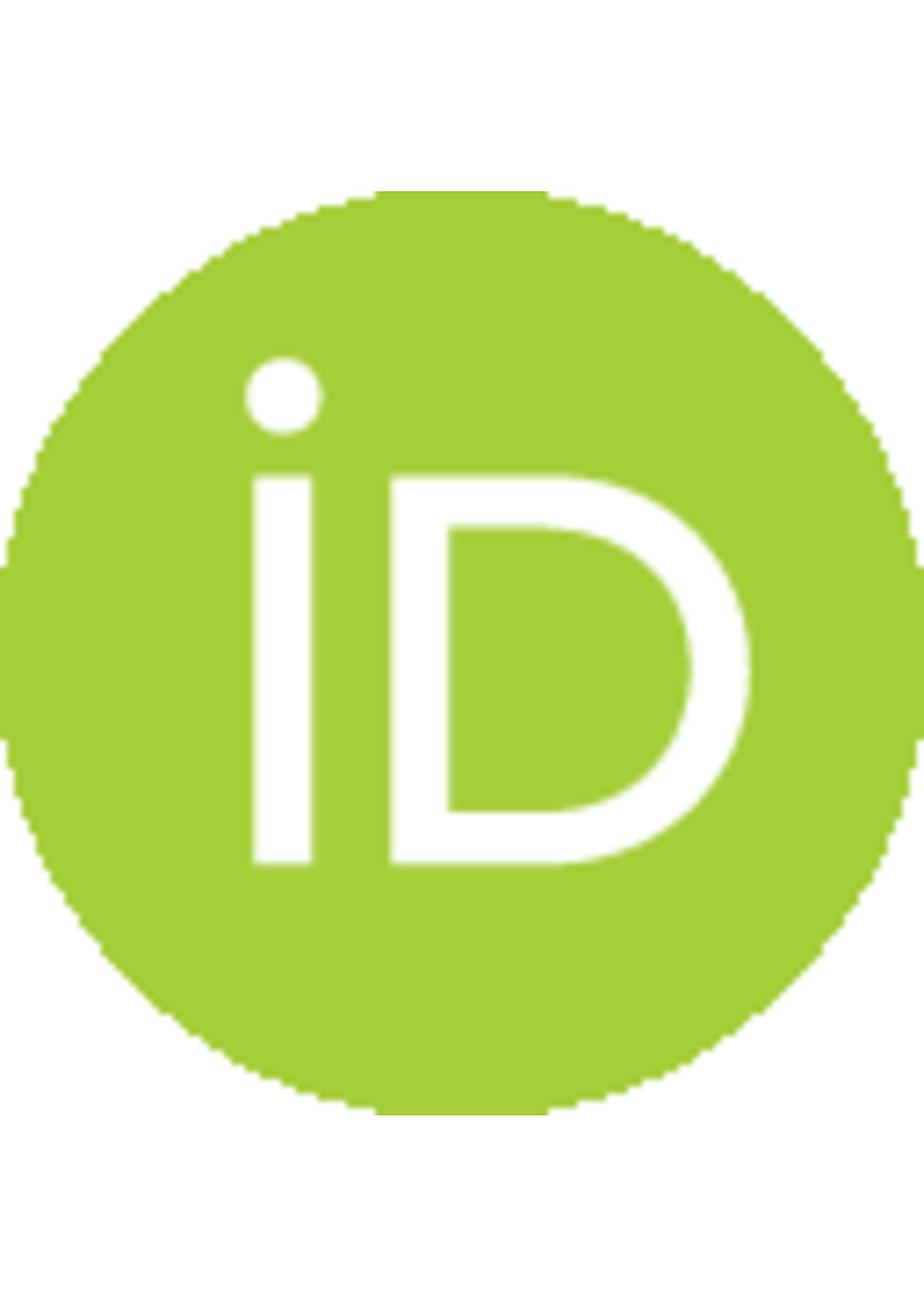}}}

\usepackage{soul}
\usepackage{xcolor}
\setstcolor{blue}
\usepackage{cancel}

\newcommand{\mathst}[1]{%
  \tikz[baseline=(X.base)]{
    \node[inner sep=0pt, outer sep=0pt] (X) {$#1$};
    \draw[blue, line width=0.5pt] (X.west) -- (X.east);
  }%
}
\usepackage{etoolbox}
\robustify{\mathst}

\usepackage[table,xcdraw]{xcolor}  
\definecolor{lightgray}{gray}{0.9} 

\def\be{\begin{equation}}
\def\ee{\end{equation}}
\def\bea{\begin{eqnarray}}
\def\eea{\end{eqnarray}}
\def\dd{{\rm d}}

\DeclareMathAlphabet{\mathpzc}{OT1}{pzc}{m}{it}
\title{\textbf{Turnover detection using the power spectrum and bispectrum}}

\author[a]{Yolanda Dube,}
\author[a]{Bikash R. Dinda,\!\orcid{0000-0001-5432-667X}}
\author[b,a]{Sheean Jolicoeur,}
\author[a,c]{\\ Roy Maartens\orcid{0000-0001-9050-5894}}

\affiliation[a]{Department of Physics \& Astronomy, University of the Western Cape, Cape Town 7535, South Africa}
\affiliation[b]{VIT Mauritius, Uniciti International Education Hub, 72448, Pierrefonds, Mauritius}
\affiliation[c]{National Institute for Theoretical \& Computational Sciences, Cape Town 7535, South Africa}

\emailAdd{yolymatics007@gmail.com}
\emailAdd{bikashrdinda@gmail.com}
\emailAdd{jolicoeursheean@gmail.com}
\emailAdd{rmaartens@uwc.ac.za}

\abstract{
The turnover at the peak of the Fourier matter power spectrum encodes a fundamental signature of matter-radiation equality in the early Universe. This delivers a potential standard ruler, independent of baryon acoustic oscillations and therefore able to break parameter degeneracies and improve precision. Furthermore, the turnover scale is independent of redshift and clustering bias, allowing for stacking of the signals from redshift bins. In practice, the very large scale of the turnover means that sample variance and systematics are serious impediments to its detection. Detections of the turnover and measurements of its scale have been made in the WiggleZ, eBOSS, Quaia, and DESI surveys. Upcoming surveys should improve the detection significance and reduce errors on the turnover scale. We use MCMC forecasts for turnover detection in a spectroscopic Euclid-like survey and a futuristic MegaMapper-like survey. In addition to the power spectrum, we include the signal from the bispectrum in equilateral configurations. 
These surveys are forecast to detect the turnover at $\sim\! 6\sigma$ (Euclid-like) and $\sim\! 15\sigma$ (MegaMapper-like), with precision on the turnover scale of $\sim\! 4\%$ and $\sim\! 2\%$. The inclusion of the bispectrum delivers a modest improvement of $\sim\! 10-17\%$ in the constraints on the turnover scale.
}

\begin{document}
\maketitle
\flushbottom

\hrule
\vspace{3pt}
\hrule
\vspace{8pt}
{\small\noindent\tableofcontents}
\vspace{8pt}
\hrule
\vspace{3pt}
\hrule
\vspace{15pt}

\section{Introduction}

The matter power spectrum $P_m(k,z)$ is a fundamental tool in cosmology that describes the clustering of matter across different scales. For the standard $\Lambda$CDM model of cosmology, it has a prominent large-scale feature corresponding to the peak in $P_m(k,z)$. The turnover that occurs at this peak is imprinted by the physics of the matter-radiation transition.  
Detecting and measuring this feature is of great scientific value for several reasons ~\cite{Blake:2004tr,prada2011measuring,poole2013wigglez,Philcox:2020xbv,Pryer:2021cut,Farren:2021grl,Philcox:2022sgj,Cunnington:2022ryj,Bahr-Kalus:2023ebd, Alonso:2024emk,DESI:2025euz}. Firstly, the turnover scale $k_0$ provides a standard ruler set by well-understood early-Universe physics, independent of the BAO (baryon acoustic oscillation) sound horizon. A precise measurement of $k_0$ can constrain the comoving Hubble horizon at equality and thus the physical matter density, $\Omega_{m0} h^2$. In combination with distance data (e.g. {type Ia} supernovae), the turnover can even yield an alternative route to measure the Hubble constant $H_0$, helping to address the long-standing $H_0$ discrepancy, or to constrain the effective number of relativistic species $N_{\rm eff}$ without relying on  CMB (cosmic microwave background) or BAO calibrations.

The transition at equality from growth ($P_m\sim k^{n_s}$) to decay ($P_m\sim k^{n_s-4}$), where $n_s$ is the primordial spectral index measured by Planck \cite{Planck:2018vyg}, imprints the peak or turnover in $P_m$ approximately at
\begin{align}
  k_{\rm eq}= a_{\rm eq}\,H_{\rm eq} = H_0\sqrt{2\Omega_{m0}}\, \big(1+z_{\text{eq}}\big)^{1/2} \,, 
  \label{eq:keq}
\end{align}
where we used the Friedmann equation.

The observed turnover scale $k_0$ is not exactly equal to the theoretical equality scale $k_{\rm eq}$ due to several physical effects. The baryonic component introduces additional complexity through the coupling with photons via Thomson scattering, which affects the growth of perturbations differently than pure cold dark matter~\cite{Hu:1996cq,Eisenstein:1998hr}. Additionally, the finite thickness of the last scattering surface and the details of recombination modify the effective transfer function~\cite{Hu:1995en}. An empirical relationship between the turnover and equality scales follows from detailed numerical calculations \cite{prada2011measuring}:
\begin{align}
\frac{k_0}{1\,h\,\mathrm{Mpc}^{-1}}
&= \left(\frac{0.194}{\omega_{b0}^{0.321}}\right)
\left[
\frac{k_{\rm eq}}{1\,h\,\mathrm{Mpc}^{-1}}
\right]^{\,0.685 - 0.121\log(\omega_{b0})}\,,
\label{eq:k0_vs_keq}
\end{align}
where $\omega_{b0} = \Omega_{b0} h^2$ is the physical baryon density and $\log=\log_{10}$. This empirical relation captures the weak but non-negligible dependence of the turnover scale on the baryon density, arising from the baryon-photon coupling that affects the growth of perturbations on scales comparable to the sound horizon at recombination.
The logarithmic correction term in \autoref{eq:k0_vs_keq} reflects the fact that higher baryon densities lead to more efficient Silk damping on small scales, which slightly shifts the effective turnover position \cite{Eisenstein:1998hr}. For the Planck 2018 cosmological parameter $\omega_b = 0.0224$ \cite{Planck:2018vyg}, this relation predicts $k_0 \approx 0.85 \, k_{\text{eq}}$, indicating that the observed turnover occurs at somewhat smaller wavenumbers (larger wavelengths) than the naive equality scale.

Observationally detecting the turnover is challenging since the turnover occurs at ultra-large scales, requiring surveys of enormous volume and careful control of systematics. Cosmic variance on these scales is large, and many Fourier modes must be sampled to robustly identify the turnover. 
Several detections and measurements of the turnover have been achieved to date. The first tentative detection of the turnover, in the WiggleZ Dark Energy Survey, was made by \cite{poole2013wigglez}. More recently, measurements of $k_0$ were made in the extended Baryon Oscillation Spectroscopic Survey (eBOSS)  quasar sample, by~\cite{Bahr-Kalus:2023ebd}, in the Quaia quasar catalog by \cite{Alonso:2024emk}, and in the Dark Energy Survey Instrument (DESI) survey, luminous red galaxy and quasar samples, by \cite{DESI:2025euz}.

Here, we investigate possible improvements from future surveys. We make MCMC forecasts
of the precision on $k_0$ that could be delivered by spectroscopic surveys similar to Euclid H$\alpha$ \cite{Euclid:2019clj} and MegaMapper Lyman-break galaxies \cite{Schlegel:2022vrv}. In addition to the power spectrum monopole used in previous work, we add the information from the monopole of the equilateral bispectrum, whose turnover is at the same scale $k_0$. In order to constrain $k_0$,  we use a model-independent fit to the power spectrum shape around the turnover, using an asymmetric parabola model, which can capture the asymmetric behavior around the turnover that follows from the physics of the matter-radiation transition.

Our fiducial model is flat $\Lambda$CDM with parameters:
$H_0 = 67.66$, $\Omega_{m0} = 0.3111$, $\Omega_{b0} = 0.049$, $n_s = 0.9665$, and $\sigma_8 = 0.8102$,
corresponding to the Planck 2018 results from CMB observations using TT, TE, EE + lowE + lensing + BAO,
as reported in the last column of Table~2 in
\cite{Planck:2018vyg}. To compute the matter power spectrum {at $z=0$}, we use the \texttt{CLASS} code.

\section{Theoretical background on the turnover scale}

The linear matter power spectrum is
\begin{equation}
    P_m(k,z) = D^2(z)P_m(k,0) \, ,
    \label{eq:matter_pm}
\end{equation}
where $D$ is the growing mode of the linear growth factor, normalized to $D(0)=1$. It is clear that the turnover is independent of redshift evolution. So we can focus on the present-day matter power spectrum, $P_m(k,0) = A\, k^{n_s} T^2(k)$, where $A$ is related to the primordial power spectrum normalization, typically expressed through $\sigma_8$ or $A_s$ \cite{Planck:2018vyg}. The matter transfer function $T(k)$ encodes the effects on density perturbations of the transition from radiation- to matter-domination, occuring at the equality epoch, $a_{\rm eq}=\Omega_{r0}/\Omega_{m0}$, corresponding to a redshift of $z_{\rm eq}\approx 3400$. This transition in the background dynamics imprints a key feature in $P_m(k,0)$ \cite{Hu:1995en,Eisenstein:1997ik,dodelson2020modern,baumann2022cosmology}: modes which are inside the comoving Hubble horizon at equality, $k>k_{\rm eq}$, are suppressed relative to the super-horizon modes, $k<k_{\rm eq}$. The super-horizon modes experience little to no processing of their primordial spectrum, so that $P_m \sim k^{n_s}$ and hence $T(k)\sim 1$. By contrast, the subhorizon modes are processed by the effects of radiation, with $T(k)\sim k^{-2}$.

\begin{figure}[!htbp]
\centering
\includegraphics[width=0.47\linewidth]{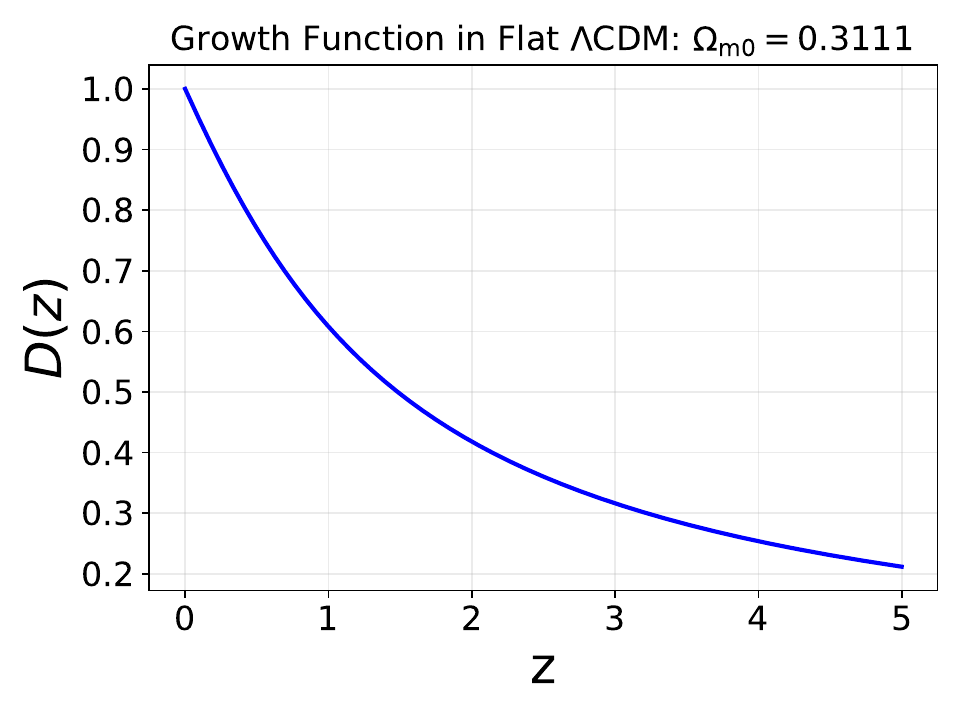}
\includegraphics[width=0.47\linewidth]{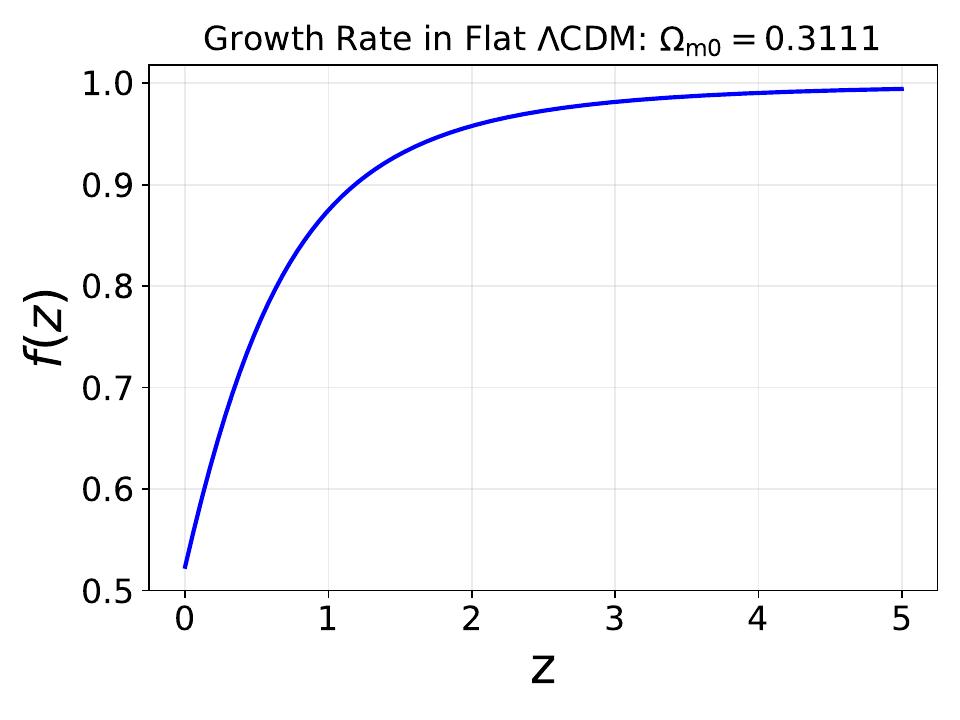}
\caption{
Growth factor ({\em left}) and growth rate ({\em right}).
}
\label{fig:growths}
\end{figure}

It is clear that $k_0$ is independent of redshift in the matter power spectrum. The same holds for the monopole of the galaxy power spectrum, since the turnover is on a very large scale -- which means that galaxy bias $b$ is scale-independent in the vicinity of the turnover. Scale-independence can be assured if the maximum mode $k_{\rm max}$ in the analysis lies well within the linear regime.
Then the galaxy monopole power is given by
\begin{equation}
P_g^{(0)}(k, z) 
= \int_{0}^{1} \dd\mu \, P_g(k, z, \mu) 
= \left[ b^2(z) + \frac{2}{3}b(z)f(z) + \frac{1}{5} f^2(z)\right] P_m(k, z) \, ,
\label{eq:monopole_redshift_space}
\end{equation}
where the linear growth rate is $f = - \dd \ln D/\dd \ln(1+z)$. At the late times relevant for  galaxy observations, we can use the analytical expressions \cite{Silveira:1994yq,Dinda:2025svh}
(see \autoref{fig:growths})
\begin{align}
D(z)& = (1+z)^{-1} \left({\, _2F_1\!\left[\frac{1}{3},1;\frac{11}{6};-{m}\right]}\right)^{-1} {\, _2F_1\! \left[\frac{1}{3},1;\frac{11}{6};-{m}(1+z)^{-3}\right]} \, ,
\label{eq:D_soln} \\
f(z)& = 1-\frac{6 m}{11 (1+z)^3} \, \frac{_2F_1\!\left[\frac{4}{3},2;\frac{17}{6};-{m}(1+z)^{-3}\right]}{ \, _2F_1\!\left[\frac{1}{3},1;\frac{11}{6};-{m}(1+z)^{-3}\right]} \, ,
\quad { \text{where} \quad m = \frac{1 - \Omega_{{m0}}}{\Omega_{{m0}}} \, . }
\label{eq:f_flcdm}
\end{align}
This allows us to compute the matter power at $z=0$ only once, using \texttt{CLASS}, and then the galaxy monopole requires only evaluations of hypergeometric functions.

\begin{figure}[!htbp]
\centering
\includegraphics[width=0.47\linewidth]{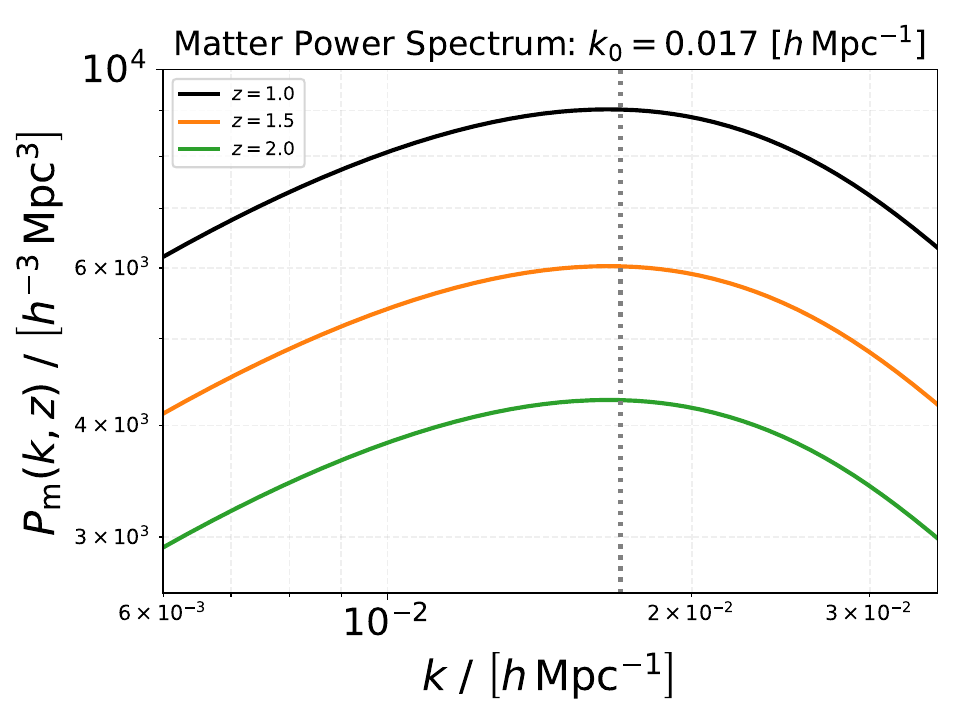}
\includegraphics[width=0.47\linewidth]{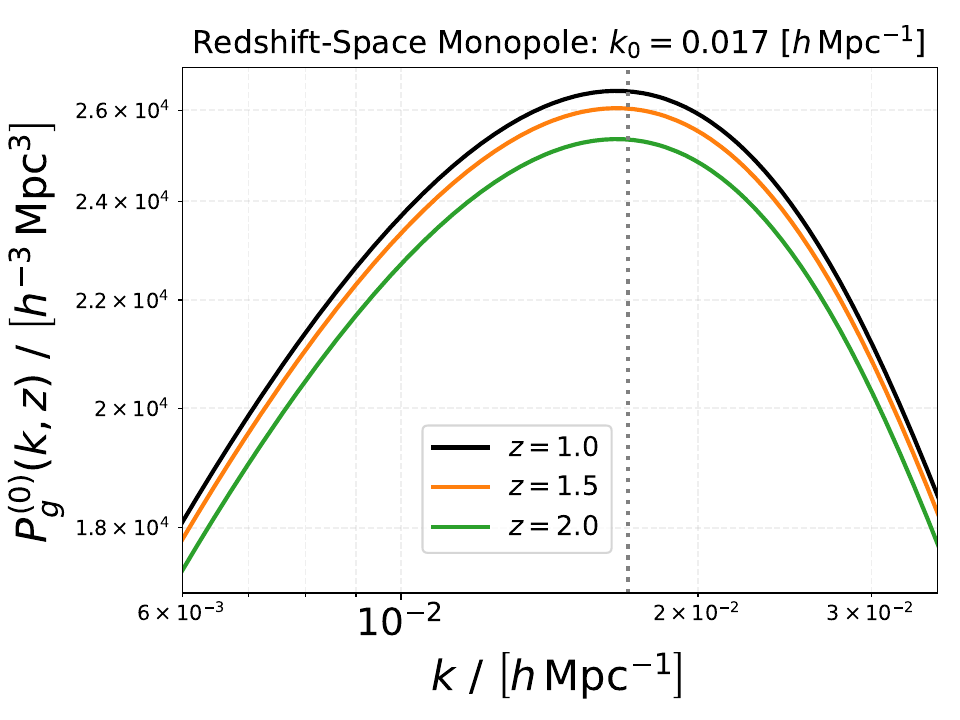}
\caption{
Matter power spectrum ({\em left}),  galaxy monopole power spectrum ({\em right}).
}
\label{fig:Power_Spectrums}
\end{figure}

\begin{figure}[!htbp]
\centering
\includegraphics[width=0.49\linewidth]{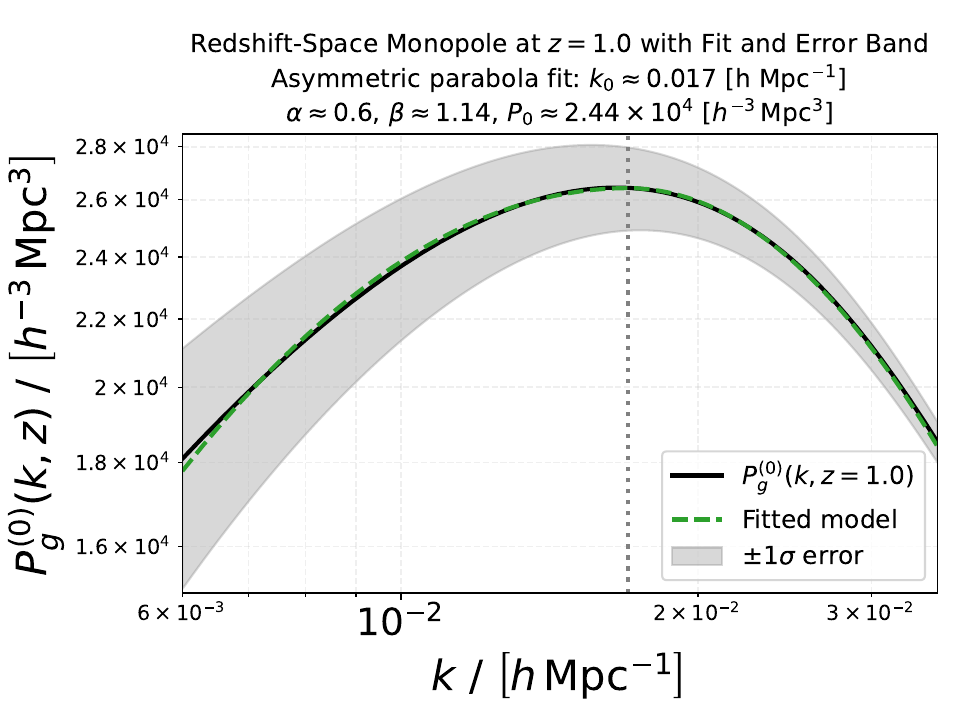}
\includegraphics[width=0.49\linewidth]{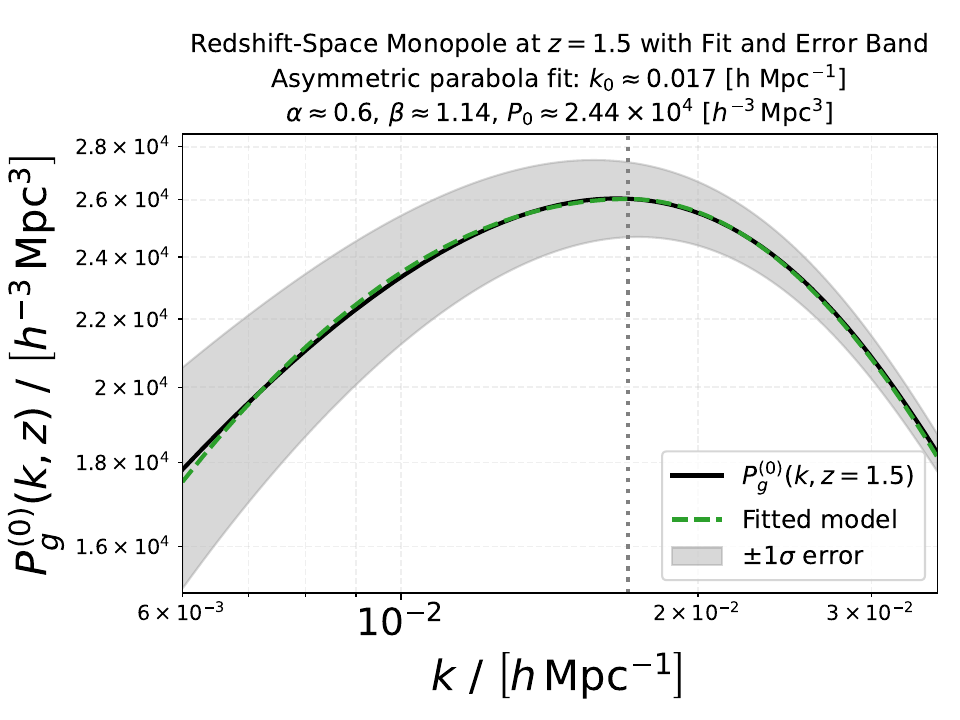}
\caption{
Monopole of redshift-space power spectrum with 1$\sigma$ errors for Euclid-like survey at redshift $z=1$ ({\em left}) and $z=1.5$ ({\em right}) computed with \texttt{CLASS} (solid black) and compared to the asymmetric parabola best-fit model in \autoref{eq:asymmetric_parabola} (dashed green and $1\sigma$ shading).
}
\label{fig:Ps0_euclid_errors}
\end{figure}

In order to detect the power spectrum turnover, we employ a widely used asymmetric parabola model  \cite{poole2013wigglez, Cunnington:2022ryj,Bahr-Kalus:2023ebd,Alonso:2024emk}, which provides a flexible parametrization that can capture the asymmetric behavior around the turnover expected from the physics of matter-radiation equality:
\begin{align}
    {P_m(k,0)} = 
    \begin{cases}
        P_0^{1 - \alpha x^2} & \text{for } k \leq k_0\,, \\[10pt]
        P_0^{1 - \beta x^2} & \text{for } k > k_0 \,,
    \end{cases}
    \quad\mbox{where}\quad x =\frac{\log\big[k/k_0\big]}{\log \big[k_0/(1\,h\,{\rm Mpc}^{-1}) \big]} \,.
    \label{eq:asymmetric_parabola}
\end{align}
It has four parameters: $P_0$ sets the peak amplitude, $k_0$ is the turnover scale, and $\alpha$, $\beta$ control the curvature below and above $k_0$, respectively (see \autoref{fig:Ps0_euclid_errors}).
Since the model is only able to fit the power spectrum accurately enough near the turnover, we impose the limits 
\begin{align}
    0.006<k/(h \text{Mpc}^{-1})<0.035\,.
\end{align}

We note that in equilateral configurations, the monopole of the galaxy bispectrum is proportional to $[P_m(k)]^2$, which has a turnover at the same scale $k_0$ as $P_m(k)$. Therefore, the equilateral bispectrum provides additional constraining power on the turnover. However, since we are dealing with large scales, we do not expect a large improvement in signal-to-noise from the bispectrum.

The tree-level bispectrum is sufficient for the large scales that we consider.
This requires second-order perturbations and, in particular, a second-order bias model.  We use the simplest, local-in-density, bias model \cite{Scoccimarro:1999ed}:
\begin{align}\label{bias}
    \delta_g = b\, \delta_{m(1)}+\frac{1}{2}b\, \delta_{m(2)} + \frac{1}{2}b_2 \big[\delta_{m(1)}\big]^2\,,
\end{align}
where $b_2$ is the quadratic bias and $\delta_{m(1,2)}$ are the first- and second-order matter density contrasts. An improved bias model would include tidal bias in \autoref{bias} \cite{Desjacques:2016bnm}. In equilateral configurations, the tidal bias is scale-independent, so that it does not affect the turnover shape or scale. Therefore, for simplicity, we neglect the tidal bias when generating mock data. 

The equilateral bispectrum is then given by \cite{Scoccimarro:1999ed}
\begin{align}
B_g^{\mathrm{equil}}(k, \mu, \nu) &= \frac{3}{448}   \bigg\{ 
63 f^4 \nu^6 \mu^2 (\mu^2 - 1)^3 \nonumber \\
& + (f \mu^2 + 4b) \Big\{ 
    f \mu^2 \Big[ 
        7 \big(4bf \mu^2  - f^2 \mu^4 + 8b^2 \big) 
        + 84b_2 + 4 f \mu^2 + 56 b 
    \Big] \nonumber \\
& + 16b (7b_2 + 4b) 
- bf \nu^2 (\mu^2 - 1) \Big[ 
    7 \big( 5 f^3 \mu^6 + 48 b^2f \mu^2 + 32b^3 \big) 
    \nonumber\\
    &+ 8 \Big( 
        56 bb_2 + 3 f \mu^2 (7b_2 - f \mu^2 + 6b) + 36 
    \Big) 
\Big] \nonumber \\
& + 3 f^2 \nu^4 (\mu^2 - 1)^2 \Big[ 
    4 (7b_2 + 3 f \mu^2 + 6b) 
    - 7 (f^2 \mu^4 - 8b^2) 
\Big] 
\bigg\}[P_m(k)]^2 \,.
\label{eq:Bs_equilateral}
\end{align}
We omit the $z$ dependence for simplicity. Here $\nu = \cos\phi$, where $\phi$ is the azimuthal angle in redshift space.
The monopole of the equilateral redshift-space galaxy bispectrum is
\begin{align}
B_g^{\mathrm{equil}(0)}(k) &=
\frac{1}{2\pi} \int_0^1 \dd\mu \int_{0}^{2\pi} \dd\phi\, B_g^{\mathrm{equil}}(k, \mu, \phi) \nonumber \\
 &= \frac{1 }{1470} \Big[
9 f^3 - 14 f^4 + 63 f^2 (14b^2 + 7 b_2 + 6b) \nonumber \\
& + 210 f (7b^3 + 14b^2 b_2 + 9b^2) + 630b^2 (7b_2 + 4b)
\Big] [P_m(k)]^2 \, .
\label{eq:Bs_monopole_final}
\end{align}

\section{{Error covariances}}
\label{sec-errors}

The auto-covariance in the redshift-space galaxy power spectrum is given by \cite{Chan:2016ehg}
\begin{equation}
\mathrm{Cov}[P_g](k, z, \mu) = \frac{4\pi^2}{V\,k^2\,\Delta k\,\Delta \mu} \left[ P_g^{\mathrm{tot}}(k,z, \mu) \right]^2 \, ,
\label{eq:covariance_ps}
\end{equation}
where 
\begin{equation}
P_g^{\mathrm{tot}}(k, z, \mu) = P_g(k, z, \mu) + \frac{1}{n_g(z)} \,.
\label{eq:ps_tot}
\end{equation}
The shot noise term is defined by the comoving background galaxy number density $n_g$, and $V$ is the comoving survey volume of the redshift bin with
width $\Delta z$, centred at $z$:
\begin{equation}
V(z) ={\frac{4\pi}{3}} f_{\rm sky} \left[ r^3(z + \Delta z/2) - r^3(z - \Delta z/2) \right] \, ,
\label{eq:comoving_volume}
\end{equation}
where $r$ is the comoving radial distance to redshift $z$ and $f_{\rm sky}$ is the  survey sky fraction.  Using linear propagation of uncertainty, we get the auto-covariance in the monopole (omitting $z$-dependence),
\begin{eqnarray}
\mathrm{Cov}[P_g^{(0)}](k) &=& \frac{2\pi^2}{V\,k^2\,\Delta k}  \int_{0}^{1} \mathrm{d}\mu \left[ P_g^{\mathrm{tot}}(k, \mu) \right]^2 \nonumber \\
&=& \frac{2\pi^2}{k^2\,n_g^2\,V\,\Delta k} \bigg[1 + 2n_g \left( b^2 + \frac{2}{3}bf + \frac{1}{5}f^2 \right) P_m(k) 
\nonumber \\&&~ 
+ n_g^2 \left( b^4 + \frac{4}{3}b^3f + \frac{6}{5}b^2f^2 + \frac{4}{7}bf^3 + \frac{1}{9}f^4 \right) [P_m(k)]^2
\bigg].
\label{eq:covariance_monopole}
\end{eqnarray}
Although the variance in the galaxy power spectrum depends on both $\Delta k$ and $\Delta \mu$, the covariance in its monopole only depends on $\Delta k= (k_{\rm max} - k_{\rm min})/{N_k}$.
Here $N_k$ is the number of $k$-bins and $k_{\rm min} = 2\pi V^{-1/3}$ is the minimum wavenumber in a given redshift bin. We set $N_k=5$, which is adequate for our constraints on the parameters.

The auto-covariance of the redshift-space equilateral bispectrum is \cite{Chan:2016ehg}
\begin{equation}
\mathrm{Cov}[B_g^{\rm equil}](k, \mu, \phi) = s^{\mathrm{equil}}\,\frac{\pi}{V}\, \frac{\left[ P_g^{\mathrm{tot}}(k, \mu) \right]^3}{k^3\,\Delta k^3\,\Delta \mu\,\Delta \phi} \, ,
\label{eq:cov_bispectrum_monopole}
\end{equation}
where $s^{\mathrm{equil}}=6$. Next, using the linear propagation of uncertainty, we get the auto-covariance in the monopole:
\begin{eqnarray}
\mathrm{Cov}[B_g^{\rm equil (0)}](k) &=& \frac{3}{2Vk^3 \Delta k^3}
\int_{0}^{1} {\mathrm{d}\mu} \int_{0}^{2\pi} \frac{\mathrm{d}\phi}{2\pi} \left[ P_g^{\mathrm{tot}}(k, \mu) \right]^3 
= \frac{3}{2Vk^3 \Delta k^3} \int_{0}^{1} {\mathrm{d}\mu} \left[ P_g^{\mathrm{tot}}(k, \mu) \right]^3 \nonumber \\
&=& \frac{3}{2n_g^3Vk^3 \Delta k^3} \Bigg[
1 + 3n_g\left( b^2 + \frac{2}{3}bf + \frac{1}{5}f^2 \right) P_m(k) \nonumber \\
&&+15n_g^2\Big(
315b^4 + 420b^3f + 378b^2f^2 + 180bf^3 + 35f^4\Big) [P_m(k)]^2 \nonumber \\
&&+1125n_g^3\Big(
3003b^6 + 6006b^5f + 9009b^4f^2 + 8580b^3f^3 \nonumber \\
&&\quad + 5005b^2f^4 + 1638bf^5 + 231f^6
\Big) [P_m(k)]^3
\Bigg] \, .
\label{eq:cov_bispectrum_monopole}
\end{eqnarray}

We compute $P_g^{(0)}(k_j, z_i)$ using \autoref{eq:monopole_redshift_space} at a particular redshift bin $z_i$ and wavenumber bin $k_j$. {We use \texttt{CLASS} to compute $P_m(k_j,0)$ and then we put this in \autoref{eq:matter_pm} to compute $P_m(k_j,z_i)$ with the help of \autoref{eq:D_soln}. Then we use  \autoref{eq:f_flcdm} and the linear bias $b(z_i)$ from a specified survey to compute the monopole of the galaxy power spectrum using \autoref{eq:monopole_redshift_space}.} Similarly, we compute $\mathrm{Cov}[P_g^{(0)}]$ at same $z_i$ and $k_j$ bins using \autoref{eq:covariance_monopole} according to a particular survey. Since there is no actual observational data, this is our fiducial or reference data with error bars. In a similar way, using \autoref{eq:monopole_redshift_space} and all other required previous equations, but replacing the present matter power spectrum by our fitting model described in \autoref{eq:asymmetric_parabola} at the same $z_i$ and $k_j$. This is our model (the `fit') to be constrained.  To constrain the model parameters, we define the individual $\chi^2$ at $z_i$ and $k_j$. We then sum over all bins to get the total $\chi^2$ of the monopole power spectrum, 
\begin{equation}
\chi^2_P = \sum_{{ij}} \frac{\left[ P_g^{(0)}(k_j, z_i)^{\mathrm{(fit)}} - P_g^{(0)}(k_j, z_i) \right]^2}{\mathrm{Cov}\left[ P_g^{(0)} \right](k_j, z_i)} \, .
\label{eq:chi2_pg_fit}
\end{equation}
We minimize this $\chi^2$ (equivalently,  maximize the likelihood, $\ln{\mathcal{L}}\sim-\chi^2/2$) to get constraints on the model parameters corresponding to Euclid- and MegaMapper-like surveys.

Similarly, we compute $\chi^2$ for the  bispectrum monopole,
\begin{equation}
\chi^2_B = \sum_{{ij}} \frac{\left[ B_g^{(0)}(k_j, z_i)^{\mathrm{(fit)}} - B_g^{(0)}(k_j, z_i) \right]^2}{\mathrm{Cov}\left[ B_g^{(0)} \right](k_j, z_i)} \, .
\label{eq:chi2_B_fit}
\end{equation}
Finally, we get the combined constraints on the model parameters by minimizing the total chi-square,
\begin{equation}
    \chi^2_{\rm tot} = \chi^2_P + \chi^2_B \, .
    \label{eq:chi2_tot}
\end{equation}

We neglect the cross-covariance between the power spectrum and bispectrum. This is expected to have a negligible impact on the signal-to-noise -- since we avoid small scales, by imposing $k/(h \text{Mpc}^{-1})<0.035$, and since we consider only equilateral configurations \cite{Chan:2016ehg,Gualdi:2020ymf,Karagiannis:2022ylq}.

To forecast the detectability of the turnover, we simulate mock observations for spectroscopic galaxy surveys similar to Euclid  \cite{Euclid:2019clj} and MegaMapper \cite{Schlegel:2022vrv}.
These surveys cover complementary redshift ranges in mapping the large-scale structure of the Universe from $z\sim 1$ to $z\sim5$, as shown in \autoref{tab:survey_comparison}. The survey number densities and clustering biases are given in \autoref{app}.

\begin{table}[htbp]
\centering
\caption{Specifications for Euclid-  and MegaMapper-like surveys}
\label{tab:survey_comparison}
\renewcommand{\arraystretch}{1.3}
\begin{tabular}{@{}lcc@{}}
\toprule
\textbf{Survey} & \textbf{Redshift Range} & \textbf{Sky Area (deg$^2$)} \\
\midrule
Euclid-like  & $0.9 < z < 1.8$ & 15,000 \\
 MegaMapper-like & $2.1 < z < 5.0$ & 20,000 \\
\bottomrule
\end{tabular}
\end{table}

To maximize the information extraction from these surveys while maintaining computational tractability, we implement a redshift bin width $\Delta z = 0.2$ bins across the full redshift range of each survey. This binning choice balances several competing factors: finer bins would better capture the redshift evolution of galaxy bias and number density, but would also increase computational cost and potentially suffer from reduced signal-to-noise in individual bins. Our $\Delta z = 0.2$ choice provides sufficient resolution to capture the key evolutionary trends while ensuring adequate galaxy counts per bin for robust statistical analysis.

\newpage
\section{Results}\label{sec-res}

\begin{figure}[!htbp]
\centering
\includegraphics[width=0.9\linewidth]{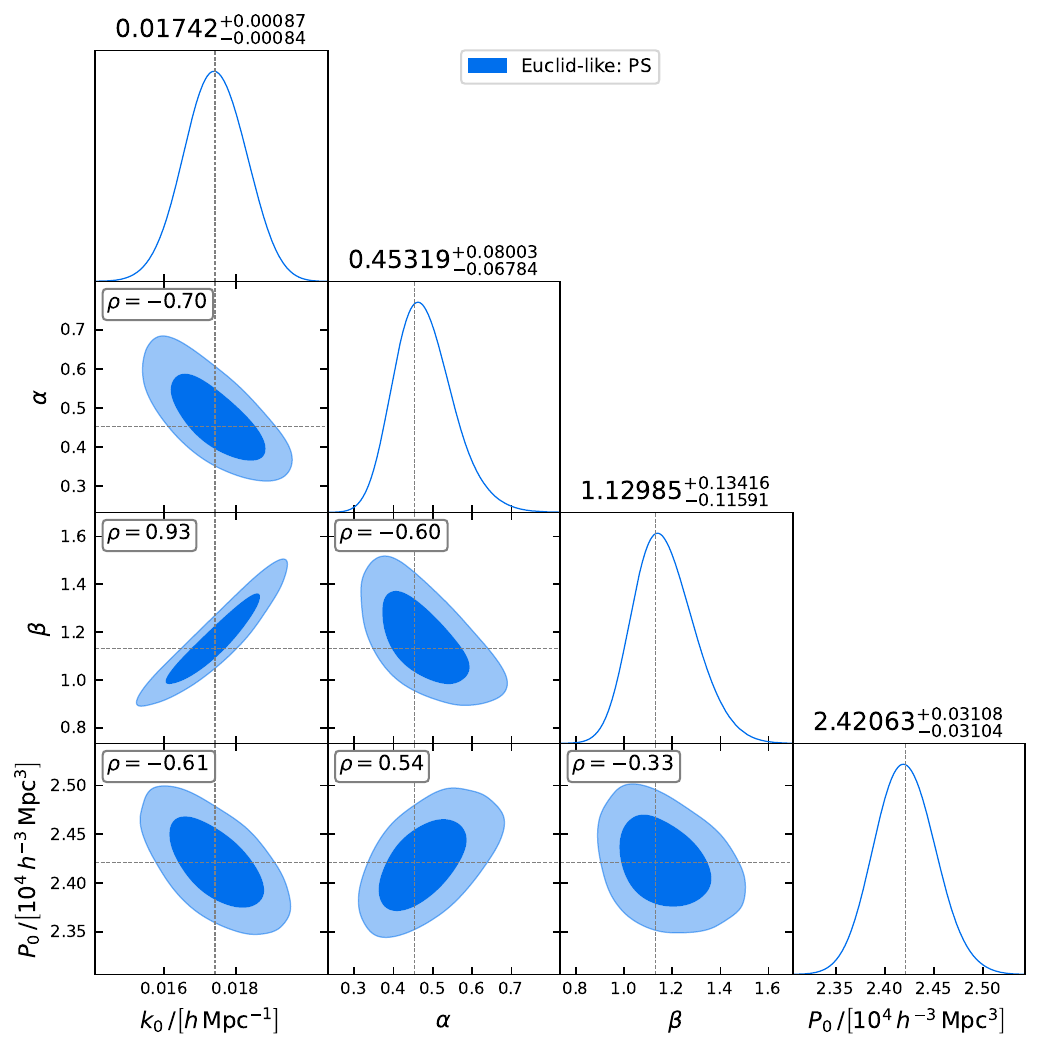}
\caption{
Triangle plot of parameter constraints from the power spectrum monopole of a Euclid-like survey (all redshift and $k$ bins combined).
}
\label{fig:Ps0_triangle}
\end{figure}

We use the \texttt{emcee} sampler \cite{Foreman-Mackey:2013rvk} to perform the chi-square minimization (maximum likelihood analysis) to infer the turnover scale and quantify parameter uncertainties. To start with, we first minimize the chi-square associated with the monopole of the galaxy power spectrum defined in \autoref{eq:chi2_pg_fit}, and consequently, we find constraints on the asymmetric parabola model parameters corresponding to the Euclid-like survey. The triangle plot is shown in \autoref{fig:Ps0_triangle}. We see that the mean values of $\alpha$ and $\beta$ are around 6$\sigma$ away from zero values. This confirms the detection of the turnover in the power spectrum. The turnover scale is $k_0/(h{\rm Mpc}^{-1})=0.01742^{+0.00087}_{-0.00084}$ (at 1$\sigma$ confidence level), corresponding to $\sim4.9\%$ precision.

\begin{figure}[!htbp]
\centering
\includegraphics[width=0.9\linewidth]{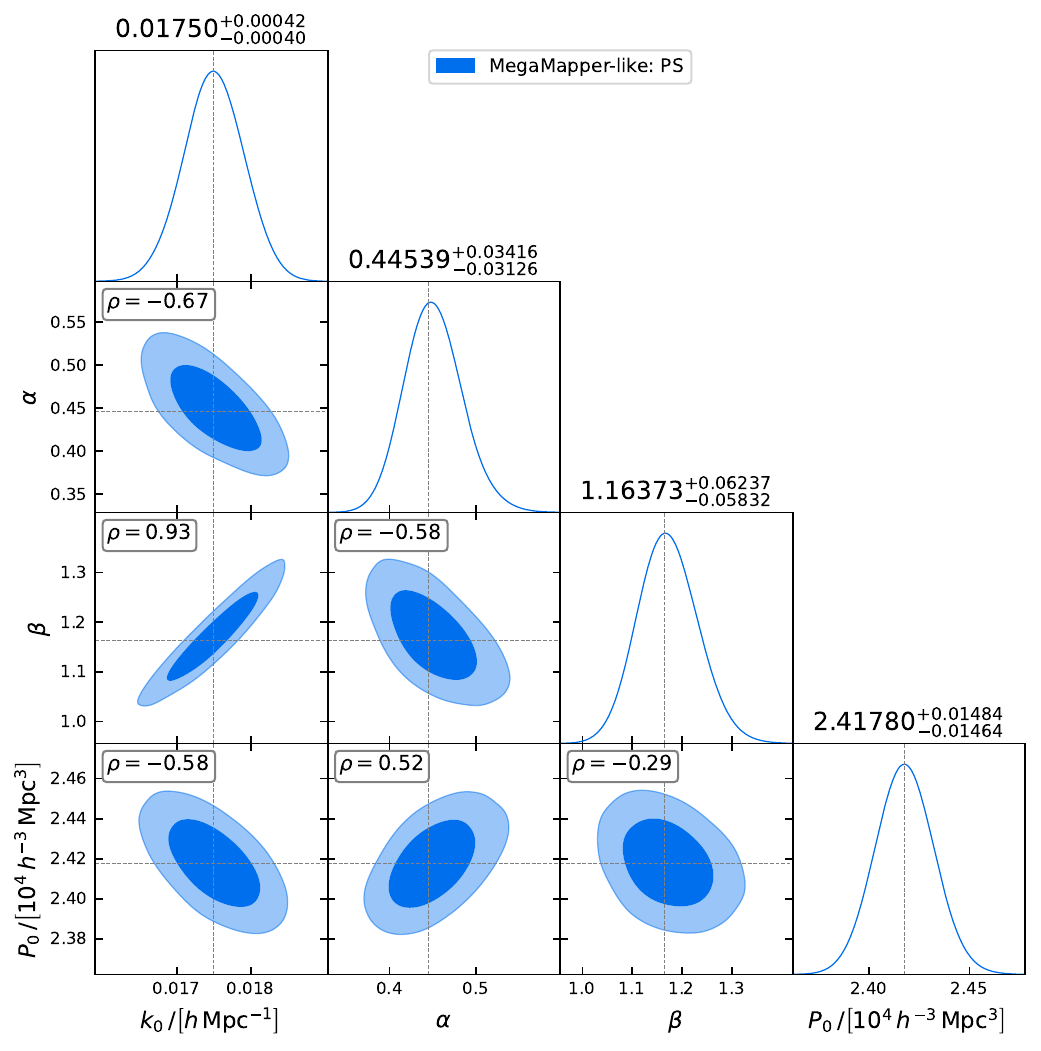}
\caption{
{As in \autoref{fig:Ps0_triangle}, for a MegaMapper-like survey.}
}
\label{fig:Ps0_triangle_MegaMapper}
\end{figure}

Similarly, we minimize the chi-square in \autoref{eq:chi2_pg_fit} according to the MegaMapper-like survey to get constraints on the model parameters using only power spectrum estimation. These are shown in \autoref{fig:Ps0_triangle_MegaMapper}. We see that turnover detection is at $\sim\! 13.5\sigma$ significance, with  $k_0/(h{\rm Mpc}^{-1})=0.01750^{+0.00042}_{-0.00040}$ (1$\sigma$), corresponding to $\sim2.3\%$ precision, less than half of the Euclid constraint. The mean values are consistent with each other.

\begin{figure}[!htbp]
\centering
\includegraphics[width=0.9\linewidth]{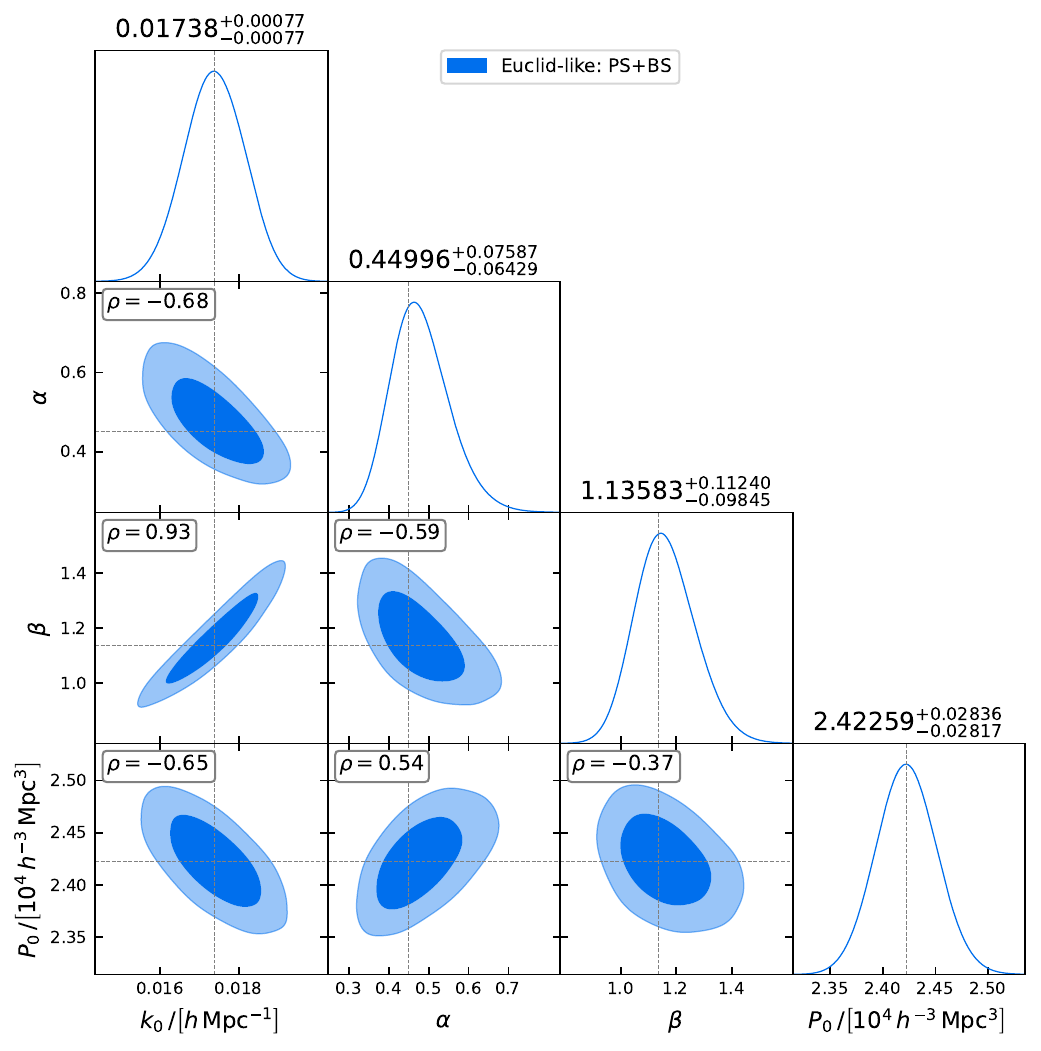}
\caption{
{Triangle plot of parameter constraints from the power spectrum + bispectrum monopoles of a Euclid-like survey (all redshift and $k$ bins combined). }
}
\label{fig:triangle_both}
\end{figure}

Next, we include constraints from the monopole of the bispectrum by minimizing the total chi-square defined in \autoref{eq:chi2_tot}. For the Euclid-like survey, the combined power spectrum + bispectrum constraints on the parameters are shown in \autoref{fig:triangle_both}. The addition of the bispectrum improves the turnover detection significance, but not by much, as expected. The improvement in the standard deviation in $k_0$ is $\sim\!10\%$.

\begin{figure}[!htbp]
\centering
\includegraphics[width=0.9\linewidth]{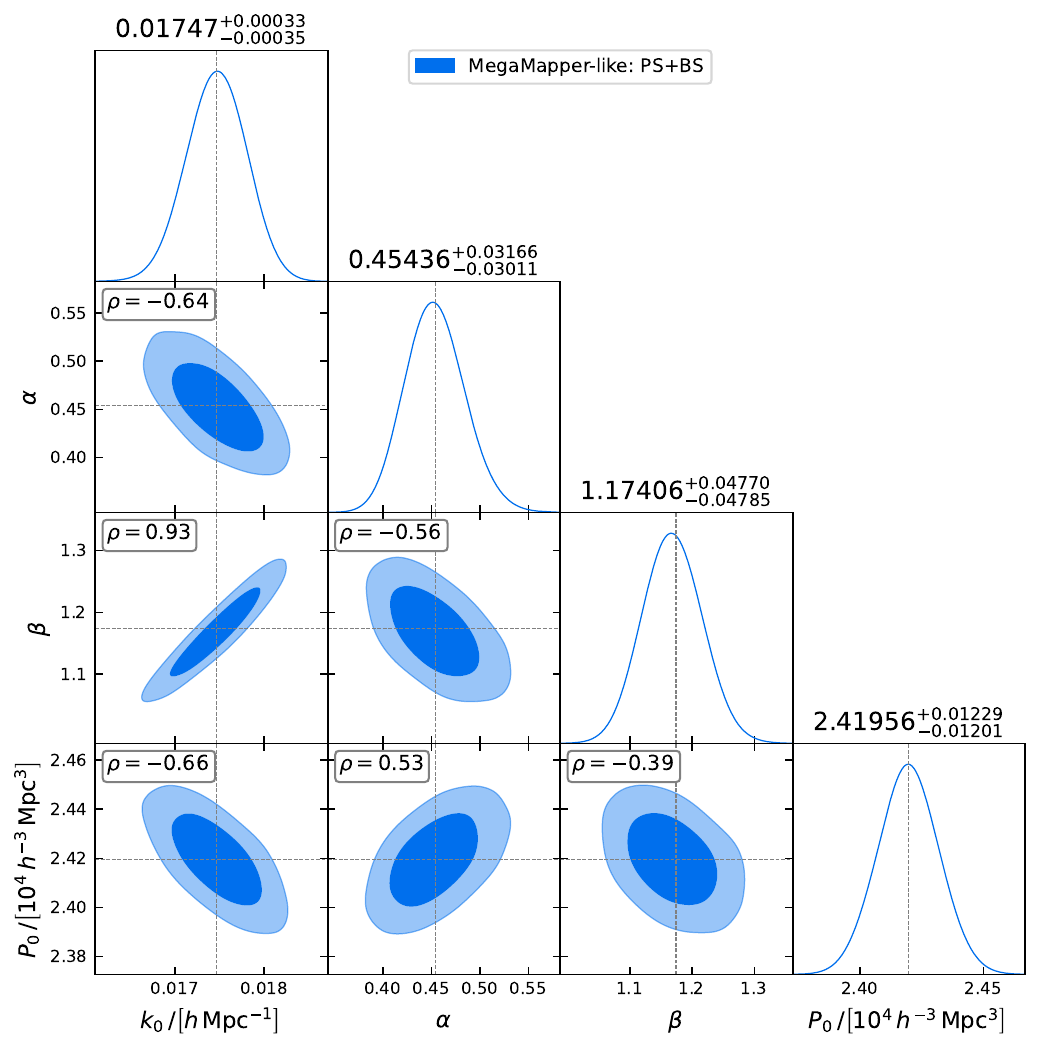}
\caption{
As in \autoref{fig:triangle_both}, for a MegaMapper-like survey.
}
\label{fig:triangle_both_MegaMapper}
\end{figure}

Finally, we repeat the power spectrum + bispectrum analysis with the Megamapper-like survey, with parameter constraints shown in \autoref{fig:triangle_both_MegaMapper}. In this case, inclusion of the bispectrum improves the detection significance to $14.6\sigma$ and tightens the precision to 1.9\%, an improvement of $\sim\! 17\%$. 

\begin{table}[htbp]
\centering
\caption{Forecast results for Euclid-  and MegaMapper-like surveys, using power spectrum (PS) and bispectrum (BS) monopoles.
}
\label{tab:forecast_results}
\renewcommand{\arraystretch}{1.3}
\begin{tabular}{@{}lccccc@{}}
\toprule
\textbf{Survey} & \textbf{Statistics} & \shortstack{\textbf{Detection} \\ \textbf{Significance}} & \textbf{$k_0 / (h\,\mathrm{Mpc}^{-1})$} & \shortstack{\textbf{Turnover} \\ \textbf{Precision}} \\
\midrule
Euclid-like & PS     & $~6.1\sigma$ & 
$0.01742\pm0.00086$ & $4.9\%$ \\
MegaMapper-like   & PS     & $13.5\sigma$ & $0.01750
\pm0.00041$ & $2.3\%$ \\
Euclid-like & PS+BS  & $~6.4\sigma$ & $0.01738 \pm 0.00077$ & 4.4$\%$ \\
MegaMapper-like   & PS+BS  & $14.6\sigma$ & $0.01747\pm0.00034$ & $1.9\%$ \\
\bottomrule
\end{tabular}
\end{table}

We summarize the main results in \autoref{tab:forecast_results}. Detection significance means the detection of an actual peak in the power spectrum. This is only possible when $\alpha$ and $\beta$ are both positive. We quantify this as the minimum of $(\alpha-0)/\Delta \alpha$,$(\beta-0)/\Delta \beta$ (in units of $\sigma$). The power spectrum falls more rapidly for $k>k_0$ compared to $k<k_0$. Therefore, we can see that in all the results, $\beta>\alpha$ (mean values). Thus detection significance is effectively $\alpha/\Delta \alpha$. Precision on the turnover is $\Delta k_0/k_0$, where we use the root-mean-square average over the slight skewness in errors. We do the same for detection significance through the symmetrization of standard deviation in $\alpha$.
\clearpage
\section{Conclusion} 
\label{sec:conclusion}

The matter power spectrum and consequently the monopole of the galaxy power spectrum have a peak value at a physical scale, given by the turnover wavenumber $k_0$. This turnover scale carries rich information about the early Universe that can be extracted at late times from galaxy counts. Specifically, the turnover scale is strongly degenerate with the radiation-matter equality scale, which in turn can provide constraints on cosmological parameters like $H_0$ and $\Omega_{m0}$, independent of BAO and other observations.

Importantly, the turnover scale in the galaxy monopole power is independent of redshift evolution, since galaxy bias and average redshift-space distortions remain scale-independent on the linear scales relevant to the turnover. This allows us to stack the signal over the redshift range of the survey.

Recent results, using samples from the {WiggleZ, eBOSS, Quaia} and DESI surveys, have already provided detections of the turnover and constraints on the turnover scale. Upcoming surveys, like those planned for Euclid or proposed for MegaMapper, can improve these constraints.

We performed an MCMC forecast for the turnover constraints from surveys similiar to Euclid  and MegaMapper, using the monopoles of the redshift space galaxy power spectrum. In each case, we stack the signals by combining the data in all redshift bins. The Euclid-like survey is forecast to constrain the turnover scale $k_0$ at $\sim\!5\%$, with a detection significance of $\sim\! 6\sigma$. The deeper and wider MegaMapper-like survey delivers more than double the precision on $k_0$, with $\sim\! 13\sigma$ detection significance.

When we include the monopole of the equilateral bispectrum, whose turnover is at the same scale as that of the power spectrum, the detection significance and the precision on the turnover scale increase appreciably, but not strongly, as expected. The constraints on $k_0$ from the combination of power spectrum and bispectrum improve by 10\% for the Euclid-like survey and by 17\% for the Megamapper-like survey.

We did not consider other bispectrum shapes since their monopoles do not give direct access to  the power spectrum turnover.  The equilateral case is special in that the bispectrum monopole is of the form ${\cal B} [P(k)]^2$, where ${\cal B}$ is scale-independent, as shown in \autoref{eq:Bs_monopole_final}. The squeezed bispectum monopole is of the form
$[{\cal B}_0+{\cal B}_2 k_L^{-2} +{\cal B}_4 k_L^{-4}] P(k_L) P(k_S)$, where ${\cal B}_a$ are scale-independent and $k_L\ll k_S$ (see \cite{Koyama:2018ttg}).
This implies that a turnover in the squeezed bispectrum monopole will not be at the turnover scale of the power spectrum monopole. The same holds for flattened (and other) configurations. (An example of the 3 cases is shown in Fig. 4 of \cite{Maartens:2020jzf}.)

There is a caveat that is relevant for upcoming surveys: in high-redshift surveys with wide sky area, there are important corrections to the power spectrum, arising from wide-separation effects (i.e. wide-angle and unequal-time) and relativistic effects \cite{Addis:2025rre}  (see also \cite{Grimm:2020ays,Castorina:2021xzs,Noorikuhani:2022bwc,Foglieni:2023xca,Guedezounme:2024pbj}). These effects come into play  for $k\lesssim k_0$, 
but they have a mostly negligible impact on the scales that we use to detect the turnover,  $0.006 \le k/(h/\text{Mpc}) \le 0.035$, and in the redshift range that we consider, $0.9\leq z \le 5$. 
However, in very high-redshift bins, the lensing corrections for the MegaMapper-like survey have a non-negligible impact on the scales close to the minimum scale $0.006h/$\,Mpc (see Fig. 2 in \cite{Addis:2025rre}). This could lead to a bias in the recovery of the true turnover scale, and we plan to investigate these corrections in future work.

Finally, we emphasize that our forecasts are optimistic, since they do not include the survey window function and observational systematics.
These introduce  problems, such as spurious density fluctuations, mode-coupling and non-diagonal covariance, on scales around  the  turnover. The measured galaxy overdensity is contaminated on very large scales by non-cosmological fluctuations due to survey and instrumental systematics, including: Galactic dust extinction and stars mistaken for galaxies; masked regions of the sky and incompleteness;
variations in observing conditions; spatially varying selection effects; imaging and redshift errors. Details on window functions and systematics relevant to turnover detection may be found in \cite{Alonso:2024emk,Bahr-Kalus:2023ebd,Pryer:2021cut}.

\vfill
\acknowledgments
YD, BRD and RM are supported by the South African Radio Astronomy Observatory and the National Research Foundation (grant no. 75415). SJ is supported by VIT Mauritius. 

\newpage
\appendix
\section{Survey properties}\label{app}

The comoving number densities and linear and quadratic biases for the two surveys are shown below, using \cite{Rossiter:2024tvi}.

\begin{table}[h]
\centering
\caption{Comoving number density and comoving volume of redshift bins for a Euclid-like survey.}
\renewcommand{\arraystretch}{1.3}
\begin{tabular}{ccc}
\hline
$z$ & $n_g / \left[10^{-3}\, h^3\, \mathrm{Mpc}^{-3}\right]$ & $V / \left[h^{-3}\, \mathrm{Gpc}^{3}\right]$ \\
\hline
0.9 & 1.56 & 3.65 \\
1.0 & 1.26 & 4.03 \\
1.1 & 1.02 & 4.36 \\
1.2 & 0.82 & 4.65 \\
1.3 & 0.67 & 4.90 \\
1.4 & 0.54 & 5.12 \\
1.5 & 0.44 & 5.30 \\
1.6 & 0.36 & 5.45 \\
1.7 & 0.29 & 5.58 \\
1.8 & 0.24 & 5.68 \\
\hline
\end{tabular}
\label{table:euclid}
\end{table}
The comoving number density and clustering biases for a Euclid-like survey can be approximately described by smooth functions of redshift $z$, valid in the range $0.9 \leq z \leq 1.8$:
\begin{align}
n_g(z) &= \Big[1.292130 - 2.774073 \,(z - 1) + 2.779457 \,(z - 1)^2 - 1.196857 \,(z - 1)^3 \Big] \nonumber\\
& \qquad \times 10^{-3} h^3\,\mathrm{Mpc}^{-3} \,, \label{eq:euclid_ng} \\
b(z) &= 0.7 \,(1 + z) \,, \label{eq:euclid_b1} \\
b_2(z) &= -0.7367 + 0.137508 \,(z - 1) + 0.149621 \,(z - 1)^2 \nonumber\\
&= -0.4132 - 0.6585 ~ b(z) + 0.3053 ~ b(z)^2 \, . \label{eq:euclid_b2}
\end{align}

\begin{table}[h]
\centering
\caption{Comoving number density and comoving volume of redshift bins for a MegaMapper-like survey.}
\renewcommand{\arraystretch}{1.3}
\begin{tabular}{ccc}
\hline
$z$ & $n_g / \left[10^{-3}\, h^3\, \mathrm{Mpc}^{-3}\right]$ & $V / \left[h^{-3}\, \mathrm{Gpc}^{3}\right]$ \\
\hline
2.1 & 2.27 & 8.09 \\
2.2 & 1.81 & 8.15 \\
2.3 & 1.42 & 8.20 \\
2.4 & 1.13 & 8.22 \\
2.5 & 0.92 & 8.23 \\
2.6 & 0.77 & 8.23 \\
2.7 & 0.67 & 8.22 \\
2.8 & 0.60 & 8.19 \\
2.9 & 0.55 & 8.16 \\
3.0 & 0.50 & 8.12 \\
3.1 & 0.46 & 8.07 \\
3.2 & 0.41 & 8.02 \\
3.3 & 0.37 & 7.96 \\
3.4 & 0.32 & 7.90 \\
3.5 & 0.28 & 7.83 \\
3.6 & 0.25 & 7.77 \\
3.7 & 0.21 & 7.70 \\
3.8 & 0.19 & 7.62 \\
3.9 & 0.17 & 7.55 \\
4.0 & 0.15 & 7.48 \\
4.1 & 0.13 & 7.40 \\
4.2 & 0.12 & 7.32 \\
4.3 & 0.11 & 7.25 \\
4.4 & 0.10 & 7.17 \\
4.5 & 0.09 & 7.09 \\
4.6 & 0.08 & 7.02 \\
4.7 & 0.07 & 6.94 \\
4.8 & 0.06 & 6.86 \\
4.9 & 0.05 & 6.79 \\
5.0 & 0.03 & 6.71 \\
\hline
\end{tabular}
\label{table:megmapper}
\end{table}

The comoving number density in \autoref{table:megmapper} and clustering biases for the MegaMapper-like survey can be approximately described by smooth functions of redshift $z$, valid in the range $2.1 \leq z \leq 5.0$:
\begin{align}
n_g(z) &= \left[-0.401657 + \frac{1.926388}{z - 1} - \frac{0.900645}{(z - 1)^2} - \frac{0.256561}{(z - 1)^3} + \frac{2.759545}{(z - 1)^4} \right] 10^{-3} h^3\,\mathrm{Mpc}^{-3} \,, \label{eq:megmapper_ng} \\
b(z) &= 0.49\, (1 + z) + 0.11\, (1 + z)^2 \,, \label{eq:megmapper_b1} \\
b_2(z) &= 0.30 - 0.79\, b(z) + 0.20\, b(z)^2 + \frac{0.12}{b(z)} \,. \label{eq:megmapper_b2}
\end{align}


\clearpage
\bibliographystyle{JHEP}
\bibliography{References}

@article{Cunnington:2022ryj,
    author = "Cunnington, Steven",
    title = "{Detecting the power spectrum turnover with H~i intensity mapping}",
    eprint = "2202.13828",
    archivePrefix = "arXiv",
    primaryClass = "astro-ph.CO",
    doi = "10.1093/mnras/stac576",
    journal = "Mon. Not. Roy. Astron. Soc.",
    volume = "512",
    number = "2",
    pages = "2408--2425",
    year = "2022"
}

@article{Scoccimarro:1999ed,
    author = "Scoccimarro, Roman and Couchman, H. M. P. and Frieman, Joshua A.",
    title = "{The Bispectrum as a Signature of Gravitational Instability in Redshift-Space}",
    eprint = "astro-ph/9808305",
    archivePrefix = "arXiv",
    reportNumber = "FERMILAB-PUB-98-254-A, CITA-98-16",
    doi = "10.1086/307220",
    journal = "Astrophys. J.",
    volume = "517",
    pages = "531--540",
    year = "1999"
}

@article{Alonso:2024emk,
    author = "Alonso, David and Hetmantsev, Oleksandr and Fabbian, Giulio and Slosar, Anze and Storey-Fisher, Kate",
    title = "{Measurement of the power spectrum turnover scale from the cross-correlation between CMB lensing and Quaia}",
    eprint = "2410.24134",
    archivePrefix = "arXiv",
    primaryClass = "astro-ph.CO",
    month = "10",
    year = "2024"
}

@article{poole2013wigglez,
  author = {Poole, Gregory B. and Blake, Chris and Parkinson, David and Brough, Sarah and Colless, Matthew and Contreras, Carlos and Couch, Warrick and Croton, Darren J. and Croom, Scott and Davis, Tamara and others},
  title = "{The WiggleZ Dark Energy Survey: probing the epoch of radiation domination using large-scale structure}",
  journal = {Mon. Not. Roy. Astron. Soc.},
  year = {2013},
  eprint = {1211.5605},
  archivePrefix = {arXiv},
  primaryClass = {astro-ph.CO}
}

@article{Bahr-Kalus:2023ebd,
    author = "Bahr-Kalus, Benedict and Parkinson, David and Mueller, Eva-Maria",
    title = "{Measurement of the matter-radiation equality scale using the extended baryon oscillation spectroscopic survey quasar sample}",
    eprint = "2302.07484",
    archivePrefix = "arXiv",
    primaryClass = "astro-ph.CO",
    doi = "10.1093/mnras/stad1867",
    journal = "Mon. Not. Roy. Astron. Soc.",
    volume = "524",
    number = "2",
    pages = "2463--2476",
    year = "2023",
}

@article{prada2011measuring,
  title={Measuring equality horizon with the zero-crossing of the galaxy correlation function},
  author={Prada, F and Klypin, A and Yepes, G and Nuza, SE and Gottloeber, S},
  eprint={1111.2889},
  archivePrefix={arXiv},
  primaryClass={astro-ph.CO},
  year={2011}
}

@book{dodelson2020modern,
  title={Modern cosmology},
  author={Dodelson, Scott and Schmidt, Fabian},
  year={2020},
  publisher={Academic press}
}

@article{Karagiannis:2022ylq,
    author = "Karagiannis, Dionysios and Maartens, Roy and Randrianjanahary, Liantsoa F.",
    title = "{Cosmological constraints from the power spectrum and bispectrum of 21cm intensity maps}",
    eprint = "2206.07747",
    archivePrefix = "arXiv",
    primaryClass = "astro-ph.CO",
    doi = "10.1088/1475-7516/2022/11/003",
    journal = "JCAP",
    volume = "11",
    pages = "003",
    year = "2022"
}

@article{Gualdi:2020ymf,
    author = "Gualdi, Davide and Verde, Licia",
    title = "{Galaxy redshift-space bispectrum: the Importance of Being Anisotropic}",
    eprint = "2003.12075",
    archivePrefix = "arXiv",
    primaryClass = "astro-ph.CO",
    doi = "10.1088/1475-7516/2020/06/041",
    journal = "JCAP",
    volume = "06",
    pages = "041",
    year = "2020"
}

@article{Planck:2018vyg,
    author = "Aghanim, N. and others",
    collaboration = "Planck",
    title = "{Planck 2018 results. VI. Cosmological parameters}",
    eprint = "1807.06209",
    archivePrefix = "arXiv",
    primaryClass = "astro-ph.CO",
    doi = "10.1051/0004-6361/201833910",
    journal = "Astron. Astrophys.",
    volume = "641",
    pages = "A6",
    year = "2020",
    note = "[Erratum: Astron.Astrophys. 652, C4 (2021)]"
}

@book{baumann2022cosmology,
  title={Cosmology},
  author={Baumann, Daniel},
  year={2022},
  publisher={Cambridge University Press}
}

@article{Guedezounme:2024pbj,
    author = "Guedezounme, S{\^e}cloka L. and Jolicoeur, Sheean and Maartens, Roy",
    title = "{Primordial non-Gaussianity {\textemdash} the effects of relativistic and wide-angle corrections to the power spectrum}",
    eprint = "2412.06553",
    archivePrefix = "arXiv",
    primaryClass = "astro-ph.CO",
    doi = "10.1088/1475-7516/2025/07/063",
    journal = "JCAP",
    volume = "07",
    pages = "063",
    year = "2025"
}

@article{Blake:2004tr,
    author = "Blake, Chris and Bridle, Sarah",
    title = "{Cosmology with photometric redshift surveys}",
    eprint = "astro-ph/0411713",
    archivePrefix = "arXiv",
    doi = "10.1111/j.1365-2966.2005.09526.x",
    journal = "Mon. Not. Roy. Astron. Soc.",
    volume = "363",
    pages = "1329--1348",
    year = "2005"
}

@article{Euclid:2019clj,
    author = "Blanchard, A. and others",
    collaboration = "Euclid",
    title = "{Euclid preparation. VII. Forecast validation for Euclid cosmological probes}",
    eprint = "1910.09273",
    archivePrefix = "arXiv",
    primaryClass = "astro-ph.CO",
    doi = "10.1051/0004-6361/202038071",
    journal = "Astron. Astrophys.",
    volume = "642",
    pages = "A191",
    year = "2020"
}

@article{Chan:2016ehg,
    author = "Chan, Kwan Chuen and Blot, Linda",
    title = "{Assessment of the Information Content of the Power Spectrum and Bispectrum}",
    eprint = "1610.06585",
    archivePrefix = "arXiv",
    primaryClass = "astro-ph.CO",
    doi = "10.1103/PhysRevD.96.023528",
    journal = "Phys. Rev. D",
    volume = "96",
    number = "2",
    pages = "023528",
    year = "2017"
}

@article{Desjacques:2016bnm,
    author = "Desjacques, Vincent and Jeong, Donghui and Schmidt, Fabian",
    title = "{Large-Scale Galaxy Bias}",
    eprint = "1611.09787",
    archivePrefix = "arXiv",
    primaryClass = "astro-ph.CO",
    doi = "10.1016/j.physrep.2017.12.002",
    journal = "Phys. Rept.",
    volume = "733",
    pages = "1--193",
    year = "2018"
}

@article{Philcox:2022sgj,
    author = "Philcox, Oliver H. E. and Farren, Gerrit S. and Sherwin, Blake D. and Baxter, Eric J. and Brout, Dillon J.",
    title = "{Determining the Hubble constant without the sound horizon: A 3.6\% constraint on H0 from galaxy surveys, CMB lensing, and supernovae}",
    eprint = "2204.02984",
    archivePrefix = "arXiv",
    primaryClass = "astro-ph.CO",
    doi = "10.1103/PhysRevD.106.063530",
    journal = "Phys. Rev. D",
    volume = "106",
    number = "6",
    pages = "063530",
    year = "2022"
}

@article{Pryer:2021cut,
    author = "Pryer, Daniel and Smith, Robert E. and Booth, Robin and Blake, Chris and Eggemeier, Alexander and Loveday, Jon",
    title = "{The galaxy power spectrum on the lightcone: deep, wide-angle redshift surveys and the turnover scale}",
    eprint = "2111.01811",
    archivePrefix = "arXiv",
    primaryClass = "astro-ph.CO",
    doi = "10.1088/1475-7516/2022/08/019",
    journal = "JCAP",
    volume = "08",
    number = "08",
    pages = "019",
    year = "2022"
}

@article{Farren:2021grl,
    author = "Farren, Gerrit S. and Philcox, Oliver H. E. and Sherwin, Blake D.",
    title = "{Determining the Hubble constant without the sound horizon: Perspectives with future galaxy surveys}",
    eprint = "2112.10749",
    archivePrefix = "arXiv",
    primaryClass = "astro-ph.CO",
    doi = "10.1103/PhysRevD.105.063503",
    journal = "Phys. Rev. D",
    volume = "105",
    number = "6",
    pages = "063503",
    year = "2022"
}

@article{Philcox:2020xbv,
    author = "Philcox, Oliver H. E. and Sherwin, Blake D. and Farren, Gerrit S. and Baxter, Eric J.",
    title = "{Determining the Hubble Constant without the Sound Horizon: Measurements from Galaxy Surveys}",
    eprint = "2008.08084",
    archivePrefix = "arXiv",
    primaryClass = "astro-ph.CO",
    doi = "10.1103/PhysRevD.103.023538",
    journal = "Phys. Rev. D",
    volume = "103",
    number = "2",
    pages = "023538",
    year = "2021"
}

@article{DESI:2025euz,
    author = "Bahr-Kalus, B. and others",
    collaboration = "DESI",
    title = "{Model-independent measurement of the matter-radiation equality scale in DESI 2024}",
    eprint = "2505.16153",
    archivePrefix = "arXiv",
    primaryClass = "astro-ph.CO",
    reportNumber = "FERMILAB-PUB-25-0369-PPD",
    doi = "10.1103/yqm1-ybbv",
    journal = "Phys. Rev. D",
    volume = "112",
    number = "6",
    pages = "063553",
    year = "2025"
}

@article{Foreman-Mackey:2013rvk,
    author = "Foreman-Mackey, Daniel and Hogg, David W. and Lang, Dustin and Goodman, Jonathan",
    title = "{emcee: The MCMC Hammer}",
    eprint = "1202.3665",
    archivePrefix = "arXiv",
    primaryClass = "astro-ph.IM",
    doi = "10.1086/670067",
    journal = "Publ. Astron. Soc. Pac.",
    volume = "125",
    pages = "306--312",
    year = "2013"
}

@article{Eisenstein:1997ik,
    author = "Eisenstein, Daniel J. and Hu, Wayne",
    title = "{Baryonic features in the matter transfer function}",
    eprint = "astro-ph/9709112",
    archivePrefix = "arXiv",
    doi = "10.1086/304893",
    journal = "Astrophys. J.",
    volume = "496",
    pages = "605",
    year = "1998"
}

@article{Grimm:2020ays,
    author = "Grimm, Nastassia and Scaccabarozzi, Fulvio and Yoo, Jaiyul and Biern, Sang Gyu and Gong, Jinn-Ouk",
    title = "{Galaxy Power Spectrum in General Relativity}",
    eprint = "2005.06484",
    archivePrefix = "arXiv",
    primaryClass = "astro-ph.CO",
    doi = "10.1088/1475-7516/2020/11/064",
    journal = "JCAP",
    volume = "11",
    pages = "064",
    year = "2020"
}

@article{Maartens:2020jzf,
    author = "Maartens, Roy and Jolicoeur, Sheean and Umeh, Obinna and De Weerd, Eline M. and Clarkson, Chris",
    title = "{Local primordial non-Gaussianity in the relativistic galaxy bispectrum}",
    eprint = "2011.13660",
    archivePrefix = "arXiv",
    primaryClass = "astro-ph.CO",
    doi = "10.1088/1475-7516/2021/04/013",
    journal = "JCAP",
    volume = "04",
    pages = "013",
    year = "2021"
}

@article{Hu:1995en,
  title={Small scale cosmological perturbations: An Analytic approach},
  author={Hu, Wayne and Sugiyama, Naoshi},
  journal={Astrophys. J.},
  volume={471},
  pages={542-570},
  year={1996},
  eprint={astro-ph/9510117}
}

@article{Hu:1996cq,
  title={The Physics of microwave background anisotropies},
  author={Hu, Wayne and Dodelson, Scott},
  journal={Ann. Rev. Astron. Astrophys.},
  volume={40},
  pages={171-216},
  year={2002},
  eprint={astro-ph/0110414}
}

@article{Eisenstein:1998hr,
  title={Baryonic features in the matter transfer function},
  author={Eisenstein, Daniel J. and Hu, Wayne},
  journal={Astrophys. J.},
  volume={496},
  pages={605-614},
  year={1998},
  eprint={astro-ph/9709112}
}

@article{Dinda:2025svh,
    author = "Dinda, Bikash R. and Maartens, Roy and Saito, Shun and Clarkson, Chris",
    title = "{Improved null tests of {\ensuremath{\Lambda}}CDM and FLRW in light of DESI DR2}",
    eprint = "2504.09681",
    archivePrefix = "arXiv",
    primaryClass = "astro-ph.CO",
    doi = "10.1088/1475-7516/2025/08/018",
    journal = "JCAP",
    volume = "08",
    pages = "018",
    year = "2025"
}

@article{Silveira:1994yq,
    author = "Silveira, V. and Waga, I.",
    title = "{Decaying Lambda cosmologies and power spectrum}",
    reportNumber = "FERMILAB-PUB-94-082-A",
    doi = "10.1103/PhysRevD.50.4890",
    journal = "Phys. Rev. D",
    volume = "50",
    pages = "4890--4894",
    year = "1994"
}

@article{Schlegel:2022vrv,
    author = "Schlegel, David J. and others",
    title = "{The MegaMapper: A Stage-5 Spectroscopic Instrument Concept for the Study of Inflation and Dark Energy}",
    eprint = "2209.04322",
    archivePrefix = "arXiv",
    primaryClass = "astro-ph.IM",
    reportNumber = "FERMILAB-FN-1195-SCD",
    month = "9",
    year = "2022"
}

@article{Rossiter:2024tvi,
    author = "Rossiter, Samantha J. and Camera, Stefano and Clarkson, Chris and Maartens, Roy",
    title = "{Decoupling local primordial non-Gaussianity from relativistic effects in the galaxy bispectrum}",
    eprint = "2407.06301",
    archivePrefix = "arXiv",
    primaryClass = "astro-ph.CO",
    doi = "10.1088/1475-7516/2025/07/055",
    journal = "JCAP",
    volume = "07",
    pages = "055",
    year = "2025"
}

@article{Castorina:2021xzs,
    author = "Castorina, Emanuele and di Dio, Enea",
    title = "{The observed galaxy power spectrum in General Relativity}",
    eprint = "2106.08857",
    archivePrefix = "arXiv",
    primaryClass = "astro-ph.CO",
    reportNumber = "CERN-TH-2021-087",
    doi = "10.1088/1475-7516/2022/01/061",
    journal = "JCAP",
    volume = "01",
    number = "01",
    pages = "061",
    year = "2022"
}

@article{Noorikuhani:2022bwc,
    author = "Noorikuhani, Milad and Scoccimarro, Roman",
    title = "{Wide-angle and relativistic effects in Fourier-space clustering statistics}",
    eprint = "2207.12383",
    archivePrefix = "arXiv",
    primaryClass = "astro-ph.CO",
    doi = "10.1103/PhysRevD.107.083528",
    journal = "Phys. Rev. D",
    volume = "107",
    number = "8",
    pages = "083528",
    year = "2023"
}

@article{Foglieni:2023xca,
    author = "Foglieni, Matteo and Pantiri, Mattia and Di Dio, Enea and Castorina, Emanuele",
    title = "{Large Scale Limit of the Observed Galaxy Power Spectrum}",
    eprint = "2303.03142",
    archivePrefix = "arXiv",
    primaryClass = "astro-ph.CO",
    reportNumber = "CERN-TH-2023-031",
    doi = "10.1103/PhysRevLett.131.111201",
    journal = "Phys. Rev. Lett.",
    volume = "131",
    number = "11",
    pages = "111201",
    year = "2023"
}

@article{Koyama:2018ttg,
    author = "Koyama, Kazuya and Umeh, Obinna and Maartens, Roy and Bertacca, Daniele",
    title = "{The observed galaxy bispectrum from single-field inflation in the squeezed limit}",
    eprint = "1805.09189",
    archivePrefix = "arXiv",
    primaryClass = "astro-ph.CO",
    doi = "10.1088/1475-7516/2018/07/050",
    journal = "JCAP",
    volume = "07",
    pages = "050",
    year = "2018"
}

@article{Addis:2025rre,
    author = "Addis, Chris and Guedezounme, S{\^e}cloka L. and Hammond, Jessie and Clarkson, Chris and Montano, Federico and Camera, Stefano and Jolicoeur, Sheean and Maartens, Roy",
    title = "{Unbiased analysis of primordial non-Gaussianity: the multipoles of the full relativistic power spectrum}",
    eprint = "2511.09466",
    archivePrefix = "arXiv",
    primaryClass = "astro-ph.CO",
    month = "11",
    year = "2025"
}

\end{document}